\documentclass[12pt]{article}

\usepackage{graphicx,amssymb,amsmath,dsfont,txfonts} 
\usepackage{bbm} 
\usepackage{mathtools} 

\usepackage[a4paper]{geometry} 
\def\dfrac#1#2{\frac{\displaystyle\strut #1}{\displaystyle\strut #2}}
\DeclareMathOperator{\tr}{tr}
\def\bra#1{\mathinner{\langle{#1}|}}
\def\ket#1{\mathinner{|{#1}\rangle}}

\begin{document}

\title{\LARGE\bf 
Modeling and simulation of a quantum thermal noise \\ on the qubit}

\author{Fran\c{c}ois {\sc Chapeau-Blondeau}, \\
    Laboratoire Angevin de Recherche en Ing\'enierie des Syst\`emes (LARIS), \\
    Universit\'e d'Angers,
    62 avenue Notre Dame du Lac, 49000 Angers, France.
}


\maketitle

\parindent=8mm \parskip=0ex

\begin{abstract}
Quantum noise or decoherence is a major factor impacting the performance of quantum 
technologies. On the qubit, an important quantum noise, often relevant in practice, is the 
thermal noise or generalized amplitude damping noise, describing the interaction with a thermal 
bath at an arbitrary temperature. A qubit thermal noise however cannot be modeled nor directly 
simulated with a few elementary Pauli operators, but instead requires specific operators. 
Our main goal here is to construct a circuit model for simulating the thermal noise from standard 
elementary qubit operators. Starting from a common quantum-operation model based on Kraus 
operators and an associated qubit-environment model, we derive a proper Stinespring dilated 
representation for the thermal noise. This dilated unitary model is then decomposed in terms of 
simple elementary qubit operators, and converted into a circuit based on elementary quantum gates. 
We arrive at our targeted simulator circuit for the thermal noise, coming with built-in easy control 
on the noise parameters. The noise simulator is then physically implemented and tested on an IBM-Q 
quantum processor. The simulator represents a useful addition to existing libraries of quantum 
circuits for quantum processors, and it offers a new tool for investigating quantum signal and 
information processing having to cope with thermal noise.
\end{abstract}

\maketitle

\section{Introduction}

{\let\thefootnote\relax\footnote{{Preprint of a paper published by {\em Fluctuation and Noise Letters},
vol.~21, 2250060, pp.~1--17 (2022). \\
https://doi.org/10.1142/S0219477522500602 }}}
At the quantum level, quantum noise or decoherence represents the 
alteration of quantum states or signals caused by their interaction with an uncontrolled environment 
\cite{Nielsen00,Haroche06,Wilde17}. Quantum noise is a major factor impacting the performance of 
quantum technologies and quantum information processing
\cite{Schleich16,Preskill18N,Ye14}. It is therefore essential to take quantum noise 
into account when designing, testing and developing quantum methodologies and devices.
In this respect, both theoretical modeling and controlled 
physical simulation are important elements for coping with quantum noise in quantum technologies. 
For the fundamental system of quantum technologies constituted by the qubit, there exist common 
noises, such as the bit-flip noise, the phase-flip noise, and the depolarizing noise, whose action 
can be conveniently modeled with a few elementary Pauli operators \cite{Nielsen00,Wilde17}. 
With such a simple and generic constitution, these noises are relatively easy to handle and can be 
rather directly simulated from elementary gates available in circuit libraries of current 
quantum processors. Another very important qubit noise is the quantum thermal noise or generalized 
amplitude damping noise \cite{Nielsen00,Chapeau15b,Khatri20}, which describes the interaction of the 
qubit with a thermal bath at an arbitrary temperature, and which is therefore frequently relevant in 
practice. However, by contrast, this noise does not possess a simple model based on the standard 
Pauli operators, but instead requires specific operators. These specific operators are not found in 
the gate libraries of standard quantum processors. As a consequence, this thermal noise model is not 
associated with a known direct circuit implementation from elementary gates. In this study, we target 
to arrive at such a simulator circuit for the qubit thermal noise, implementable with elementary 
gates available in libraries of standard quantum processors, and enabling controlled physical 
simulation.

Our starting point is a common quantum-operation model for the thermal noise based on four Kraus 
operators and an associated qubit-environment model. From these elements, we derive a proper 
Stinespring dilated model for the thermal noise. The dilated unitary representation is then 
decomposed and expressed by means of simple elementary qubit operators. This decomposition is then 
converted into a simulator circuit for the thermal noise, constituted by elementary quantum gates, 
with also built-in easy control on the parameters of the thermal noise. This quantum circuit is 
then physically implemented and tested on an IBM-Q quantum processor accessible online, to validate 
that it behaves according to the theoretical specifications of the thermal noise model.
In this way, the results deliver the targeted quantum circuit for controlled simulation of the 
qubit thermal noise on quantum processors, offering a new tool for design and test in quantum signal 
and information processing.

\section{Modeling of the qubit thermal noise} 

\subsection{Kraus operator-sum representation}

A quantum thermal noise or generalized amplitude damping noise acting on a qubit with
$2$-dimensional Hilbert space $\mathcal{H}_2$ is commonly modeled \cite{Nielsen00,Khatri20} by 
means of the four Kraus operators with matrix representation relative to the computational basis
\begin{eqnarray}
\label{4.tgad1}
\Lambda_1 &=& \sqrt{p}
\left[ \begin{array}{cc}
1 & 0  \\
0 & \sqrt{1-\gamma}
\end{array} \right] \;, \\
\label{4.tgad2}
\Lambda_2 &=& \sqrt{p}
\left[ \begin{array}{cc}
0 & \sqrt{\gamma}  \\
0 & 0
\end{array} \right] \;, \\
\label{4.tgad3}
\Lambda_3 &=& \sqrt{1-p}
\left[ \begin{array}{cc}
\sqrt{1-\gamma} & 0  \\
0               & 1  
\end{array} \right] \;, \\
\label{4.tgad4}
\Lambda_4 &=& \sqrt{1-p}
\left[ \begin{array}{cc}
0             & 0  \\
\sqrt{\gamma} & 0
\end{array} \right] \;.
\end{eqnarray}
On a qubit with density operator $\rho \in \mathcal{L}(\mathcal{H}_2)$, the thermal noise 
realizes the quantum operation
\begin{equation}
\rho \longmapsto \rho' = \mathcal{N}(\rho) = \sum_{k=1}^4 \Lambda_k \rho \Lambda_k^\dagger \;. 
\label{trKraus1}
\end{equation} 

Such a noise provides a model \cite{Nielsen00,Khatri20}, useful to quantum technologies, to describe 
the interaction of a qubit with an 
uncontrolled environment representing a thermal bath at temperature $T$. The parameter 
$\gamma \in [0, 1]$ is a damping or coupling factor which often can be expressed as a function of 
the interaction time $t$ of the qubit with the bath as $\gamma =1-e^{-t/\tau_1} $, where $\tau_1$ is 
a time constant for the interaction (such as the spin-lattice relaxation time $\tau_1$ in magnetic 
resonance). At long interaction times $t \gg \tau_1$, then $\gamma \rightarrow 1$ and the qubit 
relaxes to the equilibrium or thermalized mixed state 
$\rho_\infty= p \ket{0}\bra{0}+ (1-p) \ket{1}\bra{1}$. At equilibrium, the qubit has 
probabilities $p$ of being measured in the ground state $\ket{0}$ and $1-p$ of being measured in the 
excited state $\ket{1}$. With the energies $E_0$ and $E_1 > E_0$ respectively for the states 
$\ket{0}$ and $\ket{1}$, the equilibrium probabilities are governed by the Boltzmann distribution,
giving
\begin{equation}
p=\dfrac{\exp[-E_0/(k_BT)]}{\exp[-E_0/(k_BT)]+\exp[-E_1/(k_BT)]}
=\dfrac{1}{1+\exp[-(E_1-E_0)/(k_BT)]} \;,
\label{4.pT}
\end{equation}
and providing the connection between the temperature $T$ and the probability $p$. As $T\rightarrow 0$ 
the probability $p\rightarrow 1$ for the ground state $\ket{0}$, while at $T\rightarrow \infty$ the 
ground state $\ket{0}$ and excited state $\ket{1}$ become equiprobable with $p=1/2$. From 
Eq.~(\ref{4.pT}), when the temperature $T$ monotonically increases above $0$ up to $\infty$, the 
probability $p$ monotonically decreases from $1$ to $1/2$. 

When the input density operator has matrix representation 
$\rho=\bigl[\rho_{00}, \rho_{01} \,;\, \rho_{01}^*, 1-\rho_{00} \bigr]$ in the computational basis, 
the transformed (output) noisy density operator delivered by Eq.~(\ref{trKraus1}) follows as
\begin{equation}
\mathcal{N}(\rho) = 
\left[ \begin{array}{cc} 
{\vrule height 4mm depth 4mm width 0mm}
(1-\gamma)\rho_{00} + \gamma p & \sqrt{1-\gamma}\,\rho_{01} \\
\sqrt{1-\gamma}\,\rho_{01}^*   & 1-(1-\gamma)\rho_{00} -\gamma p  
\end{array} \right].
\label{ro_GAD}
\end{equation}

\subsection{Stinespring dilated unitary representation} \label{Stine_sec}

An alternative model for the qubit thermal noise is by means of a Stinespring dilated unitary
representation \cite{Nielsen00,Haroche06,Holevo12,Vacchini16}. 
This approach involves the introduction of a model for the 
environment and its interaction with the qubit, in order to deduce an evolution of the qubit state 
equivalent to Eq.~(\ref{trKraus1}). Such a model is especially necessary when one wants to design a 
quantum circuit so as to physically simulate, in a controlled way, the effect of the thermal noise 
on the qubit. In turn, such physical noise simulators are useful to test quantum signal or 
information processing methodologies and algorithms under controlled noise conditions.
In addition, novel quantum phenomena recently investigated and involving a non-unitary evolution of a 
qubit as in Eq.~(\ref{trKraus1}), require for their complete determination an explicit reference to a 
model for the underlying environment producing the non-unitary evolution \cite{Abbott20,Chapeau21b}.

For a principal quantum system $Q$ (here our qubit prepared in state $\rho$) with Hilbert space
$\mathcal{H}_Q$ and experiencing a non-unitary evolution $\rho \mapsto \mathcal{N}(\rho)$, a 
Stinespring dilated unitary representation introduces an environment $E$ 
with Hilbert space $\mathcal{H}_E$. The environment $E$ is prepared in a pure quantum state 
$\ket{e_0} \in \mathcal{H}_E$.
The system-environment compound $QE$ forms a closed quantum system, starting in the separable
bipartite state $\rho \otimes \ket{e_0}\bra{e_0}$, and undergoing a joint unitary evolution by the 
unitary operator $\mathsf{U}_{QE}$. The evolved bipartite state is then reduced by partial tracing
over the environment $E$, so as to obtain the reduced density operator
\begin{equation}
\rho'= \tr_E\Bigl[\mathsf{U}_{QE} \bigl( \rho \otimes \ket{e_0}\bra{e_0} \bigr) 
\mathsf{U}_{QE}^\dagger \Bigr] 
\label{2.NroQ_1}
\end{equation}
resulting for the principal system $Q$. By the Stinespring dilation theorem 
\cite{Stinespring55,Nielsen00}, any operator-sum representation as in Eq.~(\ref{trKraus1}) defining 
a valid non-unitary evolution $\rho \mapsto \rho' = \mathcal{N}(\rho)$, can be obtained by such a 
tensoring to a larger (dilated) system that is unitarily evolved and then reduced by partial 
tracing as in Eq.~(\ref{2.NroQ_1}).

For a given quantum operation $\mathcal{N}(\cdot)$ defined by a set of Kraus operators 
$\{ \Lambda_k \}_{k=1}^K$ as in Eq.~(\ref{trKraus1}) with $K=4$, the task is then to select an 
environment model with its initial state $\ket{e_0}$ and a joint unitary evolution $\mathsf{U}_{QE}$, 
so as to satisfy $\rho' = \mathcal{N}(\rho)$ in Eq.~(\ref{2.NroQ_1}).
This can be achieved in a infinite number of ways. There however exists a canonical procedure as 
follows. For any pure state $\ket{Q} \in \mathcal{H}_Q$ of the principal system $Q$, and
initial state $\ket{e_0} \in \mathcal{H}_E$ of the environment $E$, the joint unitary evolution 
$\mathsf{U}_{QE}$ is defined by
\begin{equation}
\mathsf{U}_{QE} \ket{Q} \otimes \ket{e_0} = 
\sum_{k=1}^K \Lambda_k \ket{Q} \otimes \ket{e_k} = \ket{\psi'_{QE}} \;,
\label{2.U1}
\end{equation}
where $\{\ket{e_k} \}_{k=1}^K$ is an orthonormal basis of the $K$-dimensional space 
$\mathcal{H}_{E}$ chosen for the environment $E$. 
The density operator associated with the bipartite state $\ket{\psi'_{QE}}$ of Eq.~(\ref{2.U1}) is
\begin{equation}
\rho'_{QE}=\ket{\psi'_{QE}}\bra{\psi'_{QE}} =
\sum_{k=1}^K \sum_{k'=1}^K \Lambda_{k}\ket{Q} \bra{Q} \Lambda_{k'}^\dagger
\otimes \ket{e_k} \bra{e_{k'}} \;,
\label{2.roQE}
\end{equation}
which upon partial tracing over the environment $E$ provides the reduced density operator for
the principal quantum system $Q$ as
\begin{equation}
\rho' =\tr_E(\rho'_{QE})= \sum_{k=1}^K  \Lambda_{k}\ket{Q} \bra{Q} \Lambda_{k}^\dagger \;,
\label{2.roQ}
\end{equation}
matching the targeted quantum operation defined by the Kraus operators $\{ \Lambda_k \}_{k=1}^K$.
An initial mixed state $\rho$ of the principal system $Q$ transforms in the same way, as a convex 
sum of pure states like $\ket{Q}\bra{Q}$, by linearity of Eq.~(\ref{2.roQ}).

To obtain an environment model simulating the quantum thermal noise according to this procedure, 
the $K=4$ Kraus operators $\Lambda_k$ of Eqs.~(\ref{4.tgad1})--(\ref{4.tgad4}) imply a 
$4$-dimensional environment $E$, achievable with a pair of qubits referred to the orthonormal basis 
$\{\ket{e_k} \}_{k=1}^4 =\bigl\{\ket{00}, \ket{01}, \ket{10}, \ket{11}\bigr\}$ of 
$\mathcal{H}_E \equiv \mathcal{H}_2^{\otimes 2}$. 
Then, for an arbitrary pure state $\ket{Q}=\alpha_0 \ket{0} + \alpha_1 \ket{1} \in \mathcal{H}_2$ of 
the input qubit, Eq.~(\ref{2.U1}) leads to
\begin{eqnarray}
\label{4.U_GAD1}
\mathsf{U}_{QE} \ket{Q} \otimes \ket{e_0} &=& 
\Lambda_1\ket{Q}\otimes\ket{00}+ \Lambda_2\ket{Q}\otimes\ket{01} +
\Lambda_3\ket{Q}\otimes\ket{10}+ \Lambda_4\ket{Q}\otimes\ket{11} \\
\nonumber
&=& \sqrt{p}\Bigl[ 
\alpha_0 \ket{000}+\alpha_1 \sqrt{\gamma}\ket{001}+\alpha_1\sqrt{1-\gamma}\ket{100} \Bigr] \\
\label{4.U_GAD2}
&& \mbox{} + \sqrt{1-p}\Bigl[ 
\alpha_0\sqrt{1-\gamma}\ket{010}+\alpha_1\ket{110}+\alpha_0\sqrt{\gamma}\ket{111} \Bigr] \;,
\end{eqnarray}
that must be satisfied by selecting $\mathsf{U}_{QE}$ and $\ket{e_0}$. 

For a useful noise simulator which we target here, it is important to devise an efficient control 
over the thermal noise 
parameters $(p, \gamma)$. One potential solution could be to seek to control the noise parameters 
$(p, \gamma)$ by means of the initial state $\ket{e_0} \in \mathcal{H}_2^{\otimes 2}$ of the 
two-qubit environment model, associated with a fixed $(p, \gamma)$-independent unitary 
$\mathsf{U}_{QE}$ in Eqs.~(\ref{4.U_GAD1})--(\ref{4.U_GAD2}). One could try for instance the 
separable state $\ket{e_0} =\bigl(\sqrt{p}\ket{0}+\sqrt{1-p}\ket{1}\bigr)\otimes\bigl(
\sqrt{\gamma}\ket{0}+\sqrt{1-\gamma}\ket{1}\bigr)$, or else a more involved entangled state.
Then, usually $\ket{e_0}$ will vary in the $4$-dimensional space $\mathcal{H}_2^{\otimes 2}$ when
$(p, \gamma)$ cover $[0, 1]^2$. With $\ket{Q}$ covering $\mathcal{H}_2$, the joint state 
$\ket{Q} \otimes \ket{e_0}$ will in general vary in the $8$-dimensional space 
$\mathcal{H}_2^{\otimes 3}$. However, Eq.~(\ref{4.U_GAD2}) shows that the three-qubit transformed 
state $\mathsf{U}_{QE} \ket{Q} \otimes \ket{e_0}$ assumes no component in the subspace spanned by 
$\bigl\{\ket{011}, \ket{101} \bigr\}$ and therefore remains in a $6$-dimensional subspace of 
$\mathcal{H}_2^{\otimes 3}$. Since the $8$-dimensional space $\mathcal{H}_2^{\otimes 3}$ cannot be 
unitarily mapped onto the $6$-dimensional subspace defined by Eq.~(\ref{4.U_GAD2}), no control of 
$(p, \gamma)$ can be generally achieved via the initial state $\ket{e_0}$ associated with a fixed 
$(p, \gamma)$-independent unitary $\mathsf{U}_{QE}$ satisfying 
Eqs.~(\ref{4.U_GAD1})--(\ref{4.U_GAD2}).

\section{A simulator circuit} \label{decompos_sec}

\begin{figure}[htb]
\centerline{\includegraphics[height=20mm]{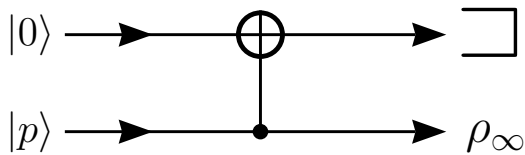}}
\caption[what appears in lof LL p177]
{Control of the parameter $p$ of the thermal noise: The Cnot gate is fed by a target qubit in 
state $\ket{0}$ and a control qubit in state $\ket{p}=\sqrt{p}\ket{0}+\sqrt{1-p}\ket{1}$. On the 
two-qubit output state $\sqrt{p}\ket{00}+\sqrt{1-p}\ket{11}$, when the target qubit is discarded, 
the control qubit is described by the mixed state 
$\rho_\infty= p \ket{0}\bra{0}+ (1-p) \ket{1}\bra{1}$.
}
\label{fig_pCnot}
\end{figure}

Instead, to guide a useful selection of the constituents $\bigl(\ket{e_0}, \mathsf{U}_{QE} \bigr)$ of 
the environment model, especially affording efficient control of the noise parameters $(p, \gamma)$,
we will make use of a model \cite{Rosati18,Khatri20} inspired from a bosonic channel for a single 
optical mode coupled to another (dissipating) optical mode at a beamsplitter with transmittivity 
$1-\gamma$ and reflectivity $\gamma$. When one of the input arm (mode) on the beamsplitter contains 
no photon, then in the incident arm (mode) a photon is transmitted unaltered with probability 
$1-\gamma$, while it is reflected into the other output arm (mode) with the complementary 
probability $\gamma$ \cite{Khatri20}. This model is known as the thermal attenuator 
\cite{Rosati18,Khatri20}, and it 
exhibits some formal equivalence with the qubit thermal noise we are interested in. In the thermal 
attenuator, the environment is represented by a two-dimensional quantum system prepared in the mixed 
state $\rho_\infty= p \ket{0}\bra{0}+ (1-p) \ket{1}\bra{1}$ determined by the equilibrium or 
thermalized state of the thermal bath. The input qubit and the two-dimensional environment interact 
through the unitary operator
\begin{equation}
\mathsf{U}_\text{th} =  
\left[ \begin{array}{cccc} 
1 & 0 & 0 & 0 \\
0 & \sqrt{1-\gamma} & \sqrt{\gamma}   & 0 \\
0 & -\sqrt{\gamma}  & \sqrt{1-\gamma} & 0 \\
0 & 0 & 0 & 1
\end{array} \right] \;,
\label{Uattenua1}
\end{equation} 
evolving the bipartite state $\rho \otimes \rho_\infty \mapsto
\mathsf{U}_\text{th} (\rho \otimes \rho_\infty) \mathsf{U}_\text{th}^\dagger$. Then it is easily 
verified that partial tracing of this transformed state over the environment produces the state 
$\mathcal{N}(\rho)$ of Eq.~(\ref{ro_GAD}). However, strictly speaking, with this model based on 
Eq.~(\ref{Uattenua1}) given in \cite{Khatri20}, we are not dealing for the qubit thermal noise 
$\mathcal{N}(\cdot)$ with a proper Stinespring dilated representation as in Eq.~(\ref{2.NroQ_1}).
The reason is that in this noise model based on Eq.~(\ref{Uattenua1}) the environment starts in a 
mixed qubit state $\rho_\infty$, while a proper Stinespring dilated representation 
\cite{Nielsen00,Haroche06,Holevo12,Vacchini16} requires an environment starting in a pure state as 
$\ket{e_0}$ in Eq.~(\ref{2.NroQ_1}). It is possible to obtain a proper Stinespring dilated 
representation as in Eq.~(\ref{2.NroQ_1}), and at the same time, for our purpose of constructing a 
simulator circuit for the thermal noise, obtain an easy control on the noise parameter $p$.
This is accomplished by introducing
a purification of the initial state $\rho_\infty$ of the environment. This can be realized by 
resorting to an additional auxiliary qubit, which is entangled to the environment qubit by a 
preparation in the pure two-qubit state $\sqrt{p}\ket{00}+\sqrt{1-p}\ket{11}$; then partial tracing 
on the auxiliary qubit places the environment qubit in the targeted mixed state 
$\rho_\infty= p \ket{0}\bra{0}+ (1-p) \ket{1}\bra{1}$. Moreover, in practice this operation is
easily accomplished by a Cnot gate fed by a target qubit in state $\ket{0}$ and a control qubit in 
state $\ket{p}=\sqrt{p}\ket{0}+\sqrt{1-p}\ket{1}$. The two qubits at the Cnot output terminate in 
the joint entangled pure state $\sqrt{p}\ket{00}+\sqrt{1-p}\ket{11}$. When the target qubit is then
discarded (non-measured), the resulting uncertainty is described by partial tracing of the joint
state over the target qubit, and the control qubit gets placed in the mixed state 
$\rho_\infty= p \ket{0}\bra{0}+ (1-p) \ket{1}\bra{1}$ as expected \cite{Nielsen00,Haroche06}. A 
quantum circuit achieving this process is depicted in Fig.~\ref{fig_pCnot}, that when input with a 
control qubit in state $\ket{p}=\sqrt{p}\ket{0}+\sqrt{1-p}\ket{1}$ will provide the targeted control 
on the parameter $p$ of the thermal noise.

The process illustrated in Fig.~\ref{fig_pCnot} restores a proper Stinespring dilated 
representation for the qubit thermal noise $\mathcal{N}(\cdot)$, with a two-qubit environment model 
prepared in the pure state $\sqrt{p}\ket{00}+\sqrt{1-p}\ket{11}=\ket{e_0}$ along with a joint 
dilated unitary evolution acting trivially on the auxiliary qubit initialized in state $\ket{0}$ 
in Fig.~\ref{fig_pCnot}, and giving in Eq.~(\ref{2.NroQ_1}) the three-qubit unitary 
$\mathsf{U}_{QE} = \mathsf{U}_\text{th} \otimes \mathrm{I}_2$.
Also, an advantage in proceeding in this way for constructing a simulator model for the qubit 
thermal noise, is that we obtain a convenient separated control of the noise parameters 
$(p, \gamma)$: the probability $p$ is controlled via an auxiliary qubit in the initial state 
$\ket{p}=\sqrt{p}\ket{0}+\sqrt{1-p}\ket{1}$ of Fig.~\ref{fig_pCnot}, while the coupling factor 
$\gamma$ is controlled via the unitary $\mathsf{U}_\text{th}$ of Eq.~(\ref{Uattenua1}).

\begin{figure}[htb]
\centerline{\includegraphics[width=140mm]{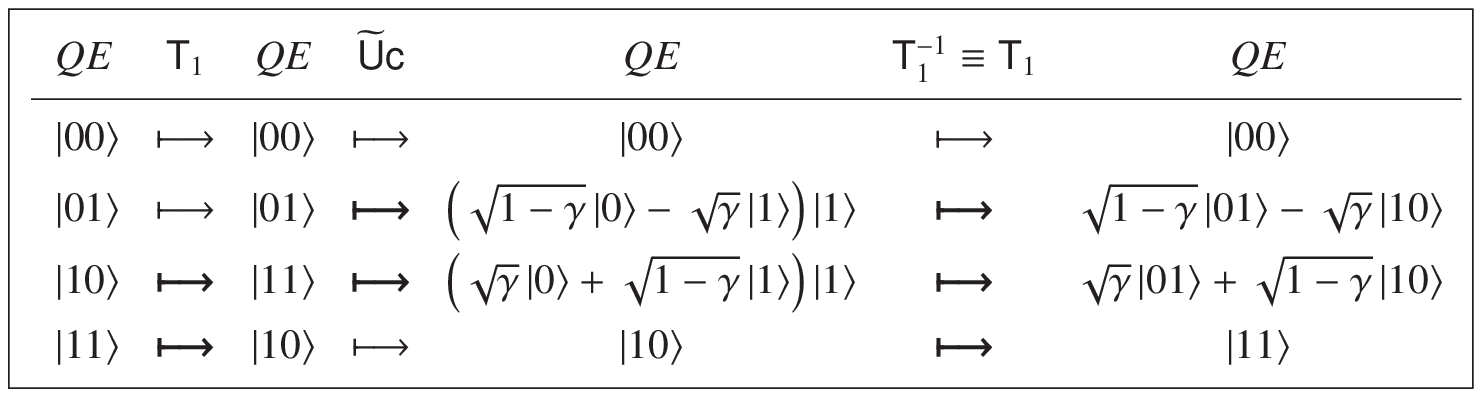}}
\caption[what appears in lof LL p177]
{The sequence of three elementary unitary transformations 
$\mathsf{T}_1$, then the controlled $\widetilde{\mathsf{U}}\mathsf{c}$ from Eq.~(\ref{Uattenua2}), 
and finally $\mathsf{T}_1^{-1} \equiv \mathsf{T}_1$, acting on the qubit pair $QE$, to realize the 
unitary transformation $\mathsf{U}_\text{th}$ of Eq.~(\ref{Uattenua1}).
}
\label{fig_trU1}
\end{figure}

It is also useful to seek a simple implementation of the two-qubit unitary $\mathsf{U}_\text{th}$ of 
Eq.~(\ref{Uattenua1}). Based on their universality property for the synthesis of quantum circuits
\cite{Barenco95,Nielsen00}, we know that the two-qubit Cnot gate complemented by one-qubit gates, 
can offer an 
implementation for $\mathsf{U}_\text{th}$ as well as for any unitary operator. Such a synthesis can be 
carried out by applying the generic procedure described in \cite{Barenco95,Nielsen00},
that we apply and report here for the first time for a simulator of the qubit thermal noise.
The procedure would start by decomposing the unitary operator into two-level unitary matrices, 
which is always feasible in principle \cite{Nielsen00} (two-level unitary matrices are unitary 
matrices that act non-trivially only on two or fewer vector components).
Here, our unitary matrix of interest $\mathsf{U}_\text{th}$ in Eq.~(\ref{Uattenua1}) is already a 
two-level unitary matrix, acting non-trivially only on the two vector components along $\ket{01}$ 
and $\ket{10}$. The two-level part of the matrix $\mathsf{U}_\text{th}$ is extracted as
\begin{equation}
\widetilde{\mathsf{U}} =  
\left[ \begin{array}{cc} 
\sqrt{1-\gamma} & \sqrt{\gamma}   \\
-\sqrt{\gamma}  & \sqrt{1-\gamma} 
\end{array} \right] \;.
\label{Uattenua2}
\end{equation} 
The two-level unitary $\mathsf{U}_\text{th}$ of Eq.~(\ref{Uattenua1}), when it acts on 
the compound $QE$ of the principal qubit $Q$ and environment qubit $E$, implements a transformation 
that can be decomposed into the sequence of three elementary unitary transformations defined in 
Fig.~\ref{fig_trU1} by their action on the computational basis of $\mathcal{H}_2^{\otimes 2}$.

\begin{figure}[htb]
\centerline{\includegraphics[height=26mm]{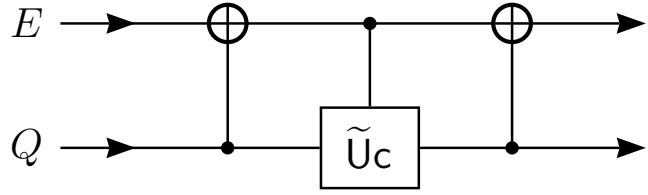}}
\caption[what appears in lof LL p177]
{Quantum circuit implementation realizing on the qubit pair $QE$ the unitary transformation 
$\mathsf{U}_\text{th}$ of Eq.~(\ref{Uattenua1}) via the decomposition of Fig.~\ref{fig_trU1},
and using a controlled version $\widetilde{\mathsf{U}}\mathsf{c}$ of the gate 
$\widetilde{\mathsf{U}}$ from Eq.~(\ref{Uattenua2}).
}
\label{fig_Ucirc}
\end{figure}

The first transformation $\mathsf{T}_1$ in Fig.~\ref{fig_trU1} rearranges the vector components 
for the qubit pair $QE$ so that the two-level transformation of Eq.~(\ref{Uattenua1}) gets applied 
only to one (the qubit $Q$) of the two quits. This transformation $\mathsf{T}_1$ amounts to a 
controlled-not on the qubit $E$ controlled by the qubit $Q$, and will therefore receive a simple 
circuit implementation with a Cnot gate. The second transformation 
$\widetilde{\mathsf{U}}\mathsf{c}$ in Fig.~\ref{fig_trU1} is the 
one-qubit unitary $\widetilde{\mathsf{U}}$ of Eq.~(\ref{Uattenua2}), acting only on qubit $Q$ 
yet under the control of qubit $E$. We will therefore have to obtain a controlled version of the
one-qubit gate $\widetilde{\mathsf{U}}$ of Eq.~(\ref{Uattenua2}). The third transformation in 
Fig.~\ref{fig_trU1} restores the initial arrangement of the vector components for the qubit pair 
$QE$, via the inverse transformation $\mathsf{T}_1^{-1} \equiv \mathsf{T}_1$ coinciding with 
$\mathsf{T}_1$. After the sequence of Fig.~\ref{fig_trU1}, the state of the qubit pair $QE$ has 
experienced the targeted unitary transformation $\mathsf{U}_\text{th}$ in Eq.~(\ref{Uattenua1}).
The sequence of Fig.~\ref{fig_trU1} can be given a simple circuit implementation depicted in 
Fig.~\ref{fig_Ucirc}.

\begin{figure}[htb]
\centerline{\includegraphics[height=26mm]{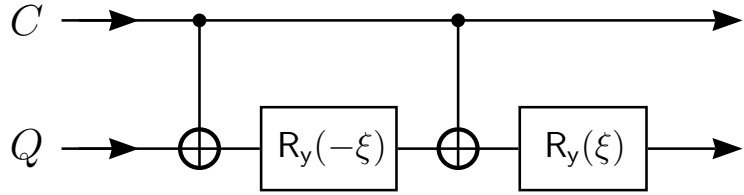}}
\caption[what appears in lof LL p177]
{Quantum circuit implementation of the controlled gate $\widetilde{\mathsf{U}}\mathsf{c}$ used in
Fig.~\ref{fig_Ucirc}, featuring two one-qubit standard $\mathsf{R_y}$ rotation gates from 
Eq.~(\ref{Ry_mat}) and two Cnot gates, and applying the transformation $\widetilde{\mathsf{U}}$ to 
the target qubit $Q$ under the control of the qubit $C$.
}
\label{fig_Ucont}
\end{figure}

We now want to work out a simple circuit implementation for the controlled gate 
$\widetilde{\mathsf{U}}\mathsf{c}$ used in Fig.~\ref{fig_Ucirc}, that is, a circuit where a 
control qubit at $\ket{1}$ applies the gate $\widetilde{\mathsf{U}}$ to the target qubit, while a 
control qubit at $\ket{0}$ leaves the target qubit unaffected. For this purpose, we follow the 
generic procedure of \cite{Barenco95,Nielsen00} to construct a controlled version of an arbitrary 
one-qubit gate. 
The unitary transformation $\widetilde{\mathsf{U}}$ of Eq.~(\ref{Uattenua2}) is a rotation in the 
space $\mathcal{H}_2$ by the angle $\xi \in [-\pi /2, 0]$ defined by $\cos(\xi)=\sqrt{1-\gamma}$ 
and $\sin(\xi)=-\sqrt{\gamma}$, that is $\xi =-\arcsin\bigl(\sqrt{\gamma} \bigr)$. It is convenient 
to introduce the standard one-qubit gate $\mathsf{R_y}(\xi )$ defined by the rotation matrix
\begin{equation}
\mathsf{R_y}(\xi )= \exp\Bigl(-i\frac{\xi}{2} \mathsf{Y} \Bigr) =
\left[ \begin{array}{lr} 
\cos(\xi /2) & -\sin(\xi /2) \\
\sin(\xi /2)  & \cos(\xi /2)   
\end{array} \right] \;,
\label{Ry_mat}
\end{equation} 
so that $\widetilde{\mathsf{U}}= \mathsf{R_y}(2\xi )$. More importantly for our purpose, we also
have $\widetilde{\mathsf{U}}= \mathsf{R_y}(\xi ) \mathsf{X} \mathsf{R_y}(-\xi ) \mathsf{X}$, with 
the standard inversion or ``not'' Pauli matrix $\mathsf{X} =[0, 1 \,;\, 1, 0]$. This decomposition 
for $\widetilde{\mathsf{U}}$ can be verified by directly performing the matrix products; it also 
results from Lemma 4.3 of \cite{Barenco95} or
Corollary 4.2 of \cite{Nielsen00} on page 176. In addition we have 
$\mathsf{R_y}(\xi ) \mathsf{R_y}(-\xi ) =\mathrm{I_2}$. It results that if we apply the control on 
$\mathsf{X}$ with a standard Cnot gate, we obtain the controlled gate 
$\widetilde{\mathsf{U}}\mathsf{c}$ that is targeted. The circuit implementation of this procedure 
is depicted in Fig.~\ref{fig_Ucont}.

\begin{figure}[htb]
\centerline{\includegraphics[height=34mm]{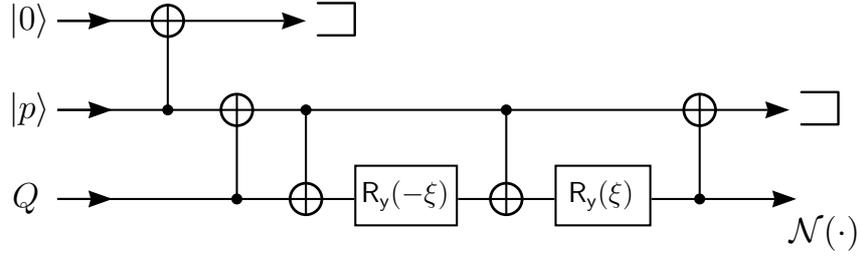}}
\caption[what appears in lof LL p177]
{Quantum circuit simulating the quantum thermal noise $\mathcal{N}(\cdot)$ defined by the quantum 
operation of Eqs.~(\ref{4.tgad1})--(\ref{trKraus1}) on the principal qubit $Q$. The input auxiliary 
qubit in state $\ket{p}=\sqrt{p}\ket{0}+\sqrt{1-p}\ket{1} \in \mathcal{H}_2$ sets the parameter $p$ 
of the thermal noise. The $2$ one-qubit rotation gates $\mathsf{R_y}(\cdot)$ with the angle 
$\xi =-\arcsin\bigl(\sqrt{\gamma} \bigr)$ set the parameter $\gamma$ of the thermal noise.
}
\label{fig_GADc}
\end{figure}

Now we can assemble the three quantum circuits of Figs.~\ref{fig_pCnot}, \ref{fig_Ucirc} and 
\ref{fig_Ucont} so as to obtain the simulator circuit for the quantum thermal noise on
the qubit with arbitrary parameters $(p, \gamma)$, as represented in Fig.~\ref{fig_GADc}.

The circuit of Fig.~\ref{fig_GADc} realizes a Stinespring dilated unitary representation according 
to the principles of Section~\ref{Stine_sec}, to simulate the qubit thermal noise 
$\mathcal{N}(\cdot)$ defined by the quantum operation of Eqs.~(\ref{4.tgad1})--(\ref{trKraus1}). The 
$4$ Kraus operators in Eqs.~(\ref{4.tgad1})--(\ref{4.tgad4}) for the qubit thermal noise determine a 
$4$-dimensional environment implemented by two auxiliary qubits in Fig.~\ref{fig_GADc}, in addition 
to the principal qubit $Q$ experiencing the thermal noise, for a total of $3$ qubits used by the 
noise simulator. The two auxiliary environment qubits are initialized in the pure separable state 
$\ket{p}\otimes\ket{0}$, with the qubit prepared in 
state $\ket{p}=\sqrt{p}\ket{0}+\sqrt{1-p}\ket{1}$ enabling one to adjust the 
parameter $p$ of the thermal noise. The quantum circuit of Fig.~\ref{fig_GADc} to simulate the qubit 
thermal noise $\mathcal{N}(\cdot)$ relies only on elementary quantum gates: $5$ two-qubit Cnot gates 
and $2$ one-qubit rotation gates $\mathsf{R_y}(\cdot)$. The angle 
$\xi =-\arcsin\bigl(\sqrt{\gamma} \bigr)$ for the rotation gates enables one to adjust the parameter 
$\gamma$ of the thermal noise. The two environment qubits in the circuit of Fig.~\ref{fig_GADc} are 
discarded (non-measured), and the circuit output marked $\mathcal{N}(\cdot)$ delivers the noisy qubit 
affected by the thermal noise with controlled parameters $(p, \gamma)$.

In the special case with $p=1$, the qubit thermal noise reduces to the amplitude damping noise 
\cite{Nielsen00}, defined by only two nonzero Kraus operators $\Lambda_1$ and $\Lambda_2$ in place 
of the four operators of Eqs.~(\ref{4.tgad1})--(\ref{4.tgad4}), and it describes the interaction
of the qubit with a thermal bath at a zero temperature $T=0$. In this special case, the control 
state $\ket{p}$ in Fig.~\ref{fig_GADc} reduces to the state $\ket{0}$, and our circuit of 
Fig.~\ref{fig_GADc} is equivalent to the quantum circuit for modeling amplitude damping given in 
Fig.~8.13 of \cite{Nielsen00}. The more general circuit of Fig.~\ref{fig_GADc} is a contribution 
to extend the noise modeling to an arbitrary temperature $T$.

\section{Physical implementation and test}

\begin{figure}[htb]
\centerline{\includegraphics[height=31mm]{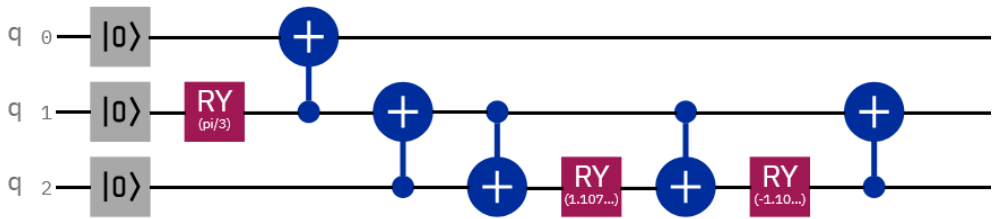}}
\caption[what appears in lof LL p177]
{Circuit layout from the front-end graphical composer interface of the IBM quantum processor, 
implementing the simulator of Fig.~\ref{fig_GADc} for the qubit thermal noise.
}
\label{fig_layout}
\end{figure}

\begin{figure}[htb]
\centerline{\includegraphics[width=55mm]{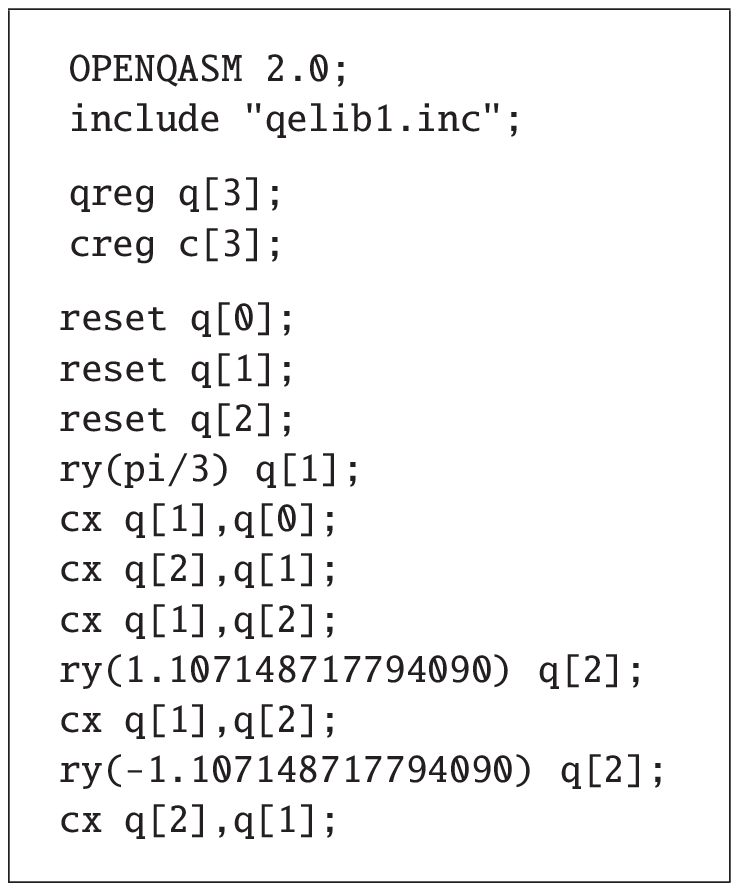}}
\caption[what appears in lof LL p177]
{Code generated by the IBM graphical composer interface to describe the quantum circuit layout 
of Fig.~\ref{fig_layout}.
}
\label{fig_code}
\end{figure}

The simulator circuit of Fig.~\ref{fig_GADc} for the qubit thermal noise has been physically 
implemented and tested on an IBM quantum processor publicly accessible online on the web
\cite{IBMq,Devitt16,Linke17,Choo18,Chen19,Shukla20,Das21}.
The quantum circuit is described via the graphical composer of the front-end interface of the 
IBM processor. The circuit description invokes a library of built-in elementary quantum gates 
available on the processor (which motivates the necessity of decomposing into elementary gates 
the quantum operation of the qubit thermal noise, as undertaken in Section~\ref{decompos_sec}).
Among standard elementary gates in the IBM quantum library are the two-qubit Cnot gate, and the 
one-qubit rotation gate $\mathsf{R_y}(\xi)$ at an arbitrary angle $\xi$. The circuit layout for the 
thermal noise simulator of Fig.~\ref{fig_GADc} is displayed in Fig.~\ref{fig_layout}, while the
code issued to describe the circuit is given in Fig.~\ref{fig_code}.

\medbreak

The input qubit state $\ket{p}=\sqrt{p}\ket{0}+\sqrt{1-p}\ket{1}$ used in the simulator of 
Fig.~\ref{fig_GADc} to set the probability $p$ of the thermal noise, is realized in the circuit of 
Fig.~\ref{fig_layout} by an additional rotation gate $\mathsf{R_y}(\xi_p)$ with angle 
$\xi_p = 2\arccos\bigl(\sqrt{p} \bigr)$, fed with the input state $\ket{0}$ and outputting 
$\ket{p}$ (acting on the qubit $q_1$ in Fig.~\ref{fig_layout}). 

\begin{figure}[htb]
\centerline{\includegraphics[height=80mm]{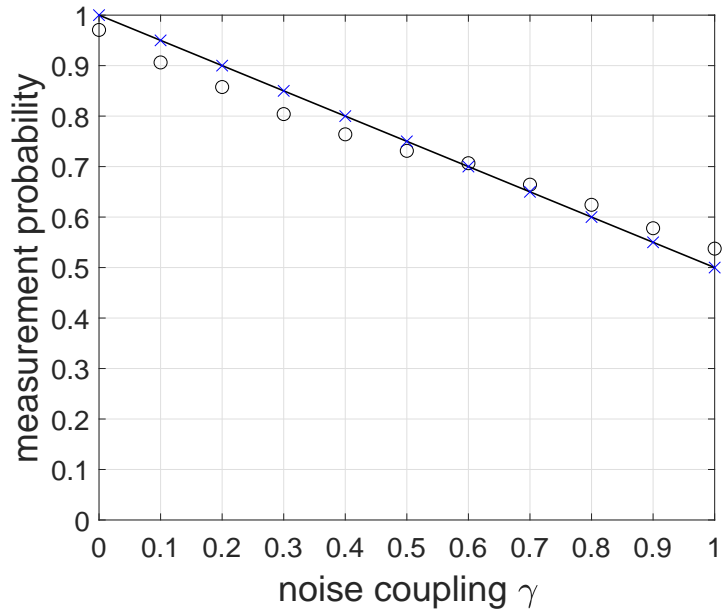}}
\caption[what appears in lof LL p177]
{For a noise probability $p=1/2$ (high temperature range) and an input qubit $Q$ in state $\ket{0}$,
the probability $\Pr\bigl\{\ket{0} \bigr\} = \bigl\langle 0 \big| 
\mathcal{N}\bigl(\ket{0}\bra{0}\bigr) \big| 0\bigr\rangle$ 
of measuring the output noisy qubit in the state $\ket{0}$, 
as a function of the coupling parameter $\gamma$ of the thermal noise.
The solid line is the theoretical model $\Pr\bigl\{\ket{0} \bigr\}=\gamma p +1-\gamma =1-\gamma /2$
from Eq.~(\ref{ro_GAD}).
The crosses ($\times$) show the theoretical value computed by the preprocessing simulator 
from the circuit layout.
The open circles ($\circ$) show the experimental evaluation.}
\label{fig_simul1}
\end{figure}

\begin{figure}[htb]
\centerline{\includegraphics[height=80mm]{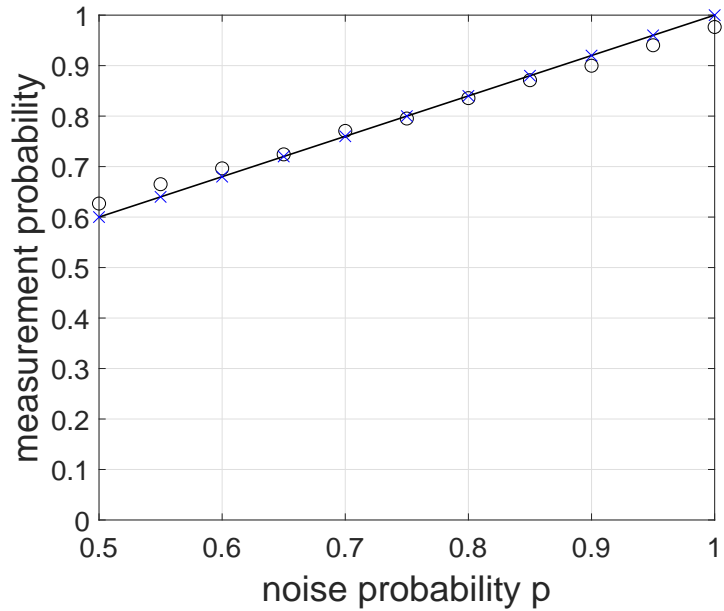}}
\caption[what appears in lof LL p177]
{For a noise coupling $\gamma=0.8$ and an input qubit $Q$ in state $\ket{0}$, 
the probability $\Pr\bigl\{\ket{0} \bigr\} = \bigl\langle 0 \big| 
\mathcal{N}\bigl(\ket{0}\bra{0}\bigr) \big| 0\bigr\rangle$ 
of measuring the output noisy qubit in the state $\ket{0}$, 
as a function of the probability $p$ of the thermal noise.
The solid line is the theoretical model $\Pr\bigl\{\ket{0} \bigr\}=\gamma p +1-\gamma =0.8p+0.2$
from Eq.~(\ref{ro_GAD}).
The crosses ($\times$) show the theoretical value computed by the preprocessing simulator
from the circuit layout.
The open circles ($\circ$) show the experimental evaluation.}
\label{fig_simul2}
\end{figure}

\begin{figure}[htb]
\centerline{\includegraphics[height=80mm]{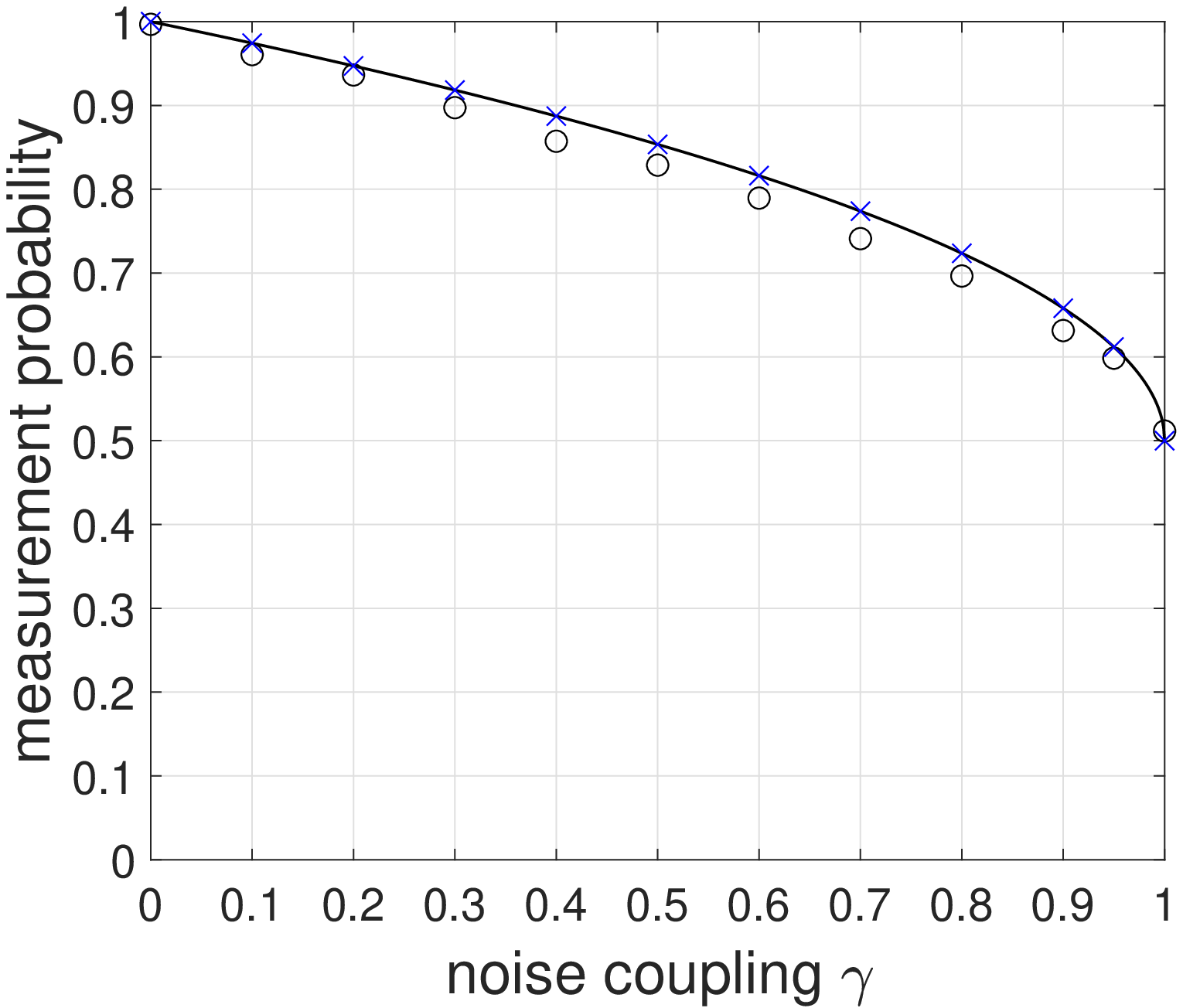}}
\caption[what appears in lof LL p177]
{For a noise probability $p=3/4$ and an input qubit $Q$ in state $\ket{+}$, 
the probability $\Pr\bigl\{\ket{+} \bigr\} = \bigl\langle + \big| 
\mathcal{N}\bigl(\ket{+}\bra{+}\bigr) \big| +\bigr\rangle$ 
of measuring the output noisy qubit in the state $\ket{+}$, 
as a function of the coupling parameter $\gamma$ of the thermal noise.
The solid line is the theoretical model $\Pr\bigl\{\ket{+} \bigr\}=\bigl(1+\sqrt{1-\gamma}\bigr)/2$
from Eq.~(\ref{ro_GAD}).
The crosses ($\times$) show the theoretical value computed by the preprocessing simulator
from the circuit layout.
The open circles ($\circ$) show the experimental evaluation.
}
\label{fig_simul3}
\end{figure}

With the noise simulator running on the quantum processor, for various states $\ket{Q}$ of the 
input qubit $Q$ (the qubit $q_2$ in Fig.~\ref{fig_layout}), quantum measurements on the output 
noisy state $\mathcal{N}\bigl(\ket{Q}\bra{Q} \bigr)$ were performed, for different configurations 
of the parameters $(p, \gamma)$ of the thermal noise. In particular, the probability 
$\Pr\bigl\{\ket{Q} \bigr\} = \bigl\langle Q \big| 
\mathcal{N}\bigl(\ket{Q}\bra{Q}\bigr) \big| Q\bigr\rangle$ of measuring the output state 
$\mathcal{N}\bigl(\ket{Q}\bra{Q} \bigr)$ in the input state $\ket{Q}$ was evaluated, which here 
equivalently represents the squared fidelity \cite{Nielsen00} of the output state 
$\mathcal{N}\bigl(\ket{Q}\bra{Q} \bigr)$ with the input state $\ket{Q}$. The experimental results 
are presented in Figs.~\ref{fig_simul1}, \ref{fig_simul2} and \ref{fig_simul3}.

\medbreak

In the tests of Figs.~\ref{fig_simul1}, \ref{fig_simul2} and \ref{fig_simul3}, for a given
state $\ket{Q}$ of the input qubit $Q$, the probabilities of the measurement outcomes on the
output noisy state $\mathcal{N}\bigl(\ket{Q}\bra{Q} \bigr)$ are theoretically predicted
from the model of Eq.~(\ref{ro_GAD}). For comparison, the same measurement probabilities are also 
theoretically evaluated, from the circuit layout of Fig.~\ref{fig_layout}, by means of the 
preprocessing simulator associated with the IBM quantum processor.
The results of Figs.~\ref{fig_simul1}, \ref{fig_simul2} and \ref{fig_simul3} always show an
excellent match between the theoretical probabilities resulting from Eq.~(\ref{ro_GAD}) and
those computed by the preprocessing simulator. This validates that the simulator circuit
designed in Fig.~\ref{fig_GADc} indeed behaves according to the specifications of the quantum
thermal noise model of Eqs.~(\ref{4.tgad1})--(\ref{trKraus1}).
Finally for comparison, the measurement probabilities have been experimentally evaluated from the 
physical implementation of the noise simulator of Figs.~\ref{fig_GADc} and \ref{fig_layout}.
The implementation has been executed on the ``ibmq\_bogota'' quantum processor from
\cite{IBMq}.
For each data point in Figs.~\ref{fig_simul1}, \ref{fig_simul2} and \ref{fig_simul3}, the run of 
the quantum circuit with the final output measurement was repeated $10^4$ times, and the 
corresponding probability value evaluated as a relative frequency.
A batch of $10^4$ runs typically took around ten seconds on the quantum processor.
The IBM quantum processors, based on superconducting qubits, have inherent imperfections.
As a result, some discrepancy is observed between the experimental and theoretical
probabilities in Figs.~\ref{fig_simul1}, \ref{fig_simul2} and \ref{fig_simul3}.
Depending on the parameter ranges, the levels of discrepancy here are quite compatible with 
those previously characterized in \cite{Linke17,Chen19,Shukla20} on former versions of the IBM 
quantum processors, or even slightly better here for processors constantly evolving and improving.
Despite these physical imperfections inherent to current experimental quantum processors 
available today, a good match is observed between theory and experiment in the results of 
Figs.~\ref{fig_simul1}, \ref{fig_simul2} and \ref{fig_simul3}, and this is obtained consistently 
over the whole range of the parameters $(p, \gamma)$ tested for the qubit thermal noise. As a 
complement, a complete experimental state tomography, as in \cite{Chen19,Gaikwad21,Zhang17}, could 
be envisaged for the output noisy qubit of the simulator circuit, and performed for every parameter 
configuration $(p, \gamma)$. This however would represent an infinite set of conditions which 
cannot be treated practically; and moreover whose detail and significance would be tied to a 
given quantum processor, while quantum processors are constantly being upgraded. For the present 
design, the tests of Figs.~\ref{fig_simul1}, \ref{fig_simul2} and \ref{fig_simul3} consolidate the 
validation of the simulator circuit and its principle, as constructed here for the qubit thermal 
noise.

\section{Conclusion}

For the modeling of the qubit thermal noise, also known as generalized amplitude damping noise,
we started from a standard quantum-operation model $\mathcal{N}(\cdot)$ in Eq.~(\ref{trKraus1}) 
based on the four specific Kraus operators of Eqs.~(\ref{4.tgad1})--(\ref{4.tgad4}) and an 
associated qubit-environment model based on Eq.~(\ref{Uattenua1}) from \cite{Rosati18,Khatri20}. 
The goal was then to exploit these modeling elements in order to construct a circuit model to 
simulate the qubit thermal noise $\mathcal{N}(\cdot)$ on a quantum processor, with easy control 
on the noise parameters $(p, \gamma)$. From Eq.~(\ref{Uattenua1}), we 
have deduced a proper Stinespring dilated model, as in Eq.~(\ref{2.NroQ_1}), arranging for 
a control of the noise parameter $p$ via an auxiliary qubit prepared in the state 
$\ket{p}=\sqrt{p}\ket{0}+\sqrt{1-p}\ket{1}$ in Figs.~\ref{fig_pCnot} and \ref{fig_GADc}.
The Stinespring dilated representation has then been decomposed in terms of a few elementary 
operators, consisting of two-qubit Cnot operators and one-qubit rotation operators.
The decomposition also arranges for an easy control of the other noise parameter, $\gamma$.
The result of the decomposition has then been converted into a three-qubit simulator circuit 
expressed with the corresponding elementary quantum gates and shown in Fig.~\ref{fig_GADc}.
The thermal noise simulator circuit obtained in Fig.~\ref{fig_GADc} is an original circuit not 
previously reported to our knowledge, and offering a useful addition to existing libraries of 
quantum circuits. This simulator circuit has then been experimentally implemented, according to the 
layout of Fig.~\ref{fig_layout}, on an IBM-Q quantum processor; and it has been experimentally 
tested and validated against the theoretical specifications of the thermal noise model. 
Through this worked-out example of a noise simulator circuit, the present study also illustrates
the generic methodology for exploiting such general-purpose quantum processors.

The present noise simulator circuit is constructed from the reference modeling approach of a 
non-unitary quantum map $\rho \mapsto \mathcal{N}(\rho)$ based on the Kraus operators in 
Eqs.~(1)--(5) and associated Stinespring dilation of Eq.~(8). Accordingly, the model for the thermal 
noise $\mathcal{N}(\cdot)$ exhibits the character of a functional-block type modeling, representing 
the action of the noise between an initial state when some processing begins and a final state when 
it ends. Such a functional-block modeling is often relevant and employed in signals and systems 
theory. In this perspective, the present simulator circuit has application in many scenarios, in 
principle anywhere a controlled thermal noise need be simulated on a signal qubit. The simulator 
can serve to design and test the performance of optimal or efficient processings having to cope 
with thermal noise, according to the noise conditions.
Quantum communication for instance offers such direction of application. Our simulator circuit 
provides a direct materialization of a quantum communication channel affected by thermal noise, and 
is well adapted to realize an input-output end-to-end simulation to investigate the noisy 
transmission across the channel, in a controlled way over the whole range of the channel 
parameters $(p, \gamma)$. Among interesting lines of investigation which can be addressed, optimal 
signaling configurations including the input signal states and their output measurements maximizing 
the input-output mutual information or achieving the information capacity, are not completely known, 
over the whole range of the channel parameters $(p, \gamma)$, especially when entangled signaling 
states of definite size are employed for communication, and in the parameter range where the channel 
is not entanglement-breaking and entangled inputs may improve over the independent-input capacity
\cite{Li-Zhen07,Holevo12,Khatri20}.
Other scenarios accessible to simulation could be the study of stochastic resonance effects,
assigning a beneficial role to quantum noise or decoherence. Stochastic resonance phenomena 
represent the (counterintuitive) possibility of enhancing the performance of some processing when the 
amount of noise is increased up to an optimal nonzero level.
For signal and information processing,
stochastic resonance has been shown to occur in the classical domain under many forms
\cite{Loerincz96,Galdi98,Chapeau99b,Gingl00,Stocks01,Kish01,Rousseau02,McDonnell02b,McDonnell08,Patel09,Delahaies13},
and in the quantum domain more recently 
\cite{Ting99,Bowen06,Wilde09k,Caruso10,Lee11,Lupo13,Gillard19}.
In the quantum domain, the possibility of such stochastic resonance phenomena has been theoretically 
shown in \cite{Chapeau15c,Gillard17,Gillard18,Gillard18b} in operations of signal detection or 
estimation from noisy quantum signals affected by thermal noise. Experimental implementation, 
validation and exploration of such phenomena in controlled noise conditions are now accessible with 
the noise simulator of Figs.~\ref{fig_GADc} and \ref{fig_layout}.
Many other possibilities can be envisaged for applying quantum noise simulators, with
controlled noise conditions, for investigations relevant to quantum signal and information 
processing.

The present study also illustrates the possibilities made available for research on quantum 
technologies by the public access, through the Internet cloud, to quantum processors, 
especially offering today an unparalleled platform for developments in quantum 
computation and quantum signal processing. 
Quantum processors, such as IBM-Q, are often exploited for quantum computation,
to implement and test quantum algorithms under the form of unitary evolutions, 
expressed by means of elementary building blocks under the form of unitary quantum gates.
The present study relates to a slightly broader perspective, illustrating that
non-unitary evolutions, such as quantum noise or decoherence,
can as well be simulated on such quantum processors, under controlled conditions.
Such possibilities are specially relevant to signal processing, which naturally
has to cope with noise while designing efficient processing, classical or quantum.
The present circuit model for quantum thermal noise offers a tool to contribute in these 
directions.


\begin{thebibliography}{10}
\newcommand{\enquote}[1]{``#1''}

\bibitem{Nielsen00}
M.~A. Nielsen and I.~L. Chuang, \emph{Quantum Computation and Quantum
  Information} (Cambridge University Press, Cambridge, 2000).

\bibitem{Haroche06}
S.~Haroche and J.-M. Raimond, \emph{Exploring the Quantum: Atoms, Cavities, and
  Photons} (Oxford University Press, Oxford, 2006).

\bibitem{Wilde17}
M.~M. Wilde, \emph{Quantum Information Theory} (Cambridge University Press,
  Cambridge, 2017).

\bibitem{Schleich16}
W.~P. Schleich, {\em et~al.}, \enquote{Quantum technology: From research to
  application}, \emph{Applied Physics B} \textbf{122} (2016) 130,1--31.

\bibitem{Preskill18N}
J.~Preskill, \enquote{Quantum computing in the {NISQ} (noisy intermediate-scale
  quantum) era and beyond}, \emph{Quantum} \textbf{2} (2018) 79,1--20.

\bibitem{Ye14}
B.~Ye and L.~Qiu, \enquote{$1/f$ noise in {I}sing quantum computers},
  \emph{Fluctuation and Noise Letters} \textbf{13} (2014) 1450006,1--8.

\bibitem{Chapeau15b}
F.~Chapeau-Blondeau, \enquote{Optimization of quantum states for signaling
  across an arbitrary qubit noise channel with minimum-error detection},
  \emph{IEEE Transactions on Information Theory} \textbf{61} (2015) 4500--4510.

\bibitem{Khatri20}
S.~Khatri, K.~Sharma and M.~M. Wilde, \enquote{Information-theoretic aspects of
  the generalized amplitude-damping channel}, \emph{Physical Review A}
  \textbf{102} (2020) 012401,1--31.

\bibitem{Holevo12}
A.~S. Holevo and V.~Giovannetti, \enquote{Quantum channels and their entropic
  characteristics}, \emph{Reports on Progress in Physics} \textbf{75} (2012)
  046001,1--30.

\bibitem{Vacchini16}
B.~Vacchini, \enquote{Quantum noise from reduced dynamics}, \emph{Fluctuation
  and Noise Letters} \textbf{15} (2016) 1640003,1--9.

\bibitem{Abbott20}
A.~A. Abbott, J.~Wechs, D.~Horsman, M.~Mhalla and C.~Branciard,
  \enquote{Communication through coherent control of quantum channels},
  \emph{Quantum} \textbf{4} (2020) 333,1--14.

\bibitem{Chapeau21b}
F.~Chapeau-Blondeau, \enquote{Quantum parameter estimation on coherently
  superposed noisy channels}, \emph{Physical Review A} \textbf{104} (2021)
  032214,1--16.

\bibitem{Stinespring55}
W.~F. Stinespring, \enquote{Positive functions on {$C^*$}-algebras},
  \emph{Proceedings of the American Mathematical Society} \textbf{6} (1955)
  211--216.

\bibitem{Rosati18}
M.~Rosati, A.~Mari and V.~Giovannetti, \enquote{Narrow bounds for the quantum
  capacity of thermal attenuators}, \emph{Nature Communications} \textbf{9}
  (2018) 4339,1--9.

\bibitem{Barenco95}
A.~Barenco, C.~H. Bennett, R.~Cleve, D.~P. DiVincenzo, N.~Margolus, P.~Shor,
  T.~Sleator, J.~A. Smolin and H.~Weinfurter, \enquote{Elementary gates for
  quantum computation}, \emph{Physical Review A} \textbf{52} (1995) 3457--3467.

\bibitem{IBMq}
\enquote{{IBM} {Q}uantum {C}omputing {P}latform},
  https://quantum-computing.ibm.com (accessed 23 Dec.\ 2021).

\bibitem{Devitt16}
S.~J. Devitt, \enquote{Performing quantum computing experiments in the cloud},
  \emph{Physical Review A} \textbf{94} (2016) 032329,1--13.

\bibitem{Linke17}
N.~M. Linke, D.~Maslov, M.~Roetteler, S.~Debnath, C.~Figgatt, K.~A. Landsman,
  K.~Wright and C.~Monroe, \enquote{Experimental comparison of two quantum
  computing architectures}, \emph{Proceedings of the National Academy of
  Sciences of the USA} \textbf{114} (2017) 3305--3310.

\bibitem{Choo18}
K.~Choo, C.~W. von Keyserlingk, N.~Regnault and T.~Neupert,
  \enquote{Measurement of the entanglement spectrum of a symmetry-protected
  topological state using the {IBM} quantum computer}, \emph{Physical Review
  Letters} \textbf{121} (2018) 086808,1--5.

\bibitem{Chen19}
Y.~Chen, M.~Farahzad, S.~Yoo and T.-C. Wei, \enquote{Detector tomography on
  {IBM} quantum computers and mitigation of an imperfect measurement},
  \emph{Physical Review A} \textbf{100} (2019) 052315,1--17.

\bibitem{Shukla20}
A.~Shukla, M.~Sisodia and A.~Pathak, \enquote{Complete characterization of the
  directly implementable quantum gates used in the {IBM} quantum processors},
  \emph{Physics Letters A} \textbf{384} (2020) 126387,1--8.

\bibitem{Das21}
S.~Das, M.~D. Rahman and M.~Majumdar, \enquote{Design of a quantum repeater
  using quantum circuits and benchmarking its performance on an {IBM} quantum
  computer}, \emph{Quantum Information Processing} \textbf{20} (2021)
  245,1--17.

\bibitem{Gaikwad21}
A.~Gaikwad, K.~Shende and K.~Dorai, \enquote{Experimental demonstration of
  optimized quantum process tomography on the {IBM} quantum experience},
  \emph{International Journal of Quantum Information} \textbf{19} (2021)
  2040004,1--15.

\bibitem{Zhang17}
J.~Zhang, K.~Li, S.~Cong and H.~Wang, \enquote{Efficient reconstruction of
  density matrices for high dimensional quantum state tomography}, \emph{Signal
  Processing} \textbf{139} (2017) 136--142.

\bibitem{Li-Zhen07}
H.~Li-Zhen and F.~Mao-Fa, \enquote{The {H}olevo capacity of a generalized
  amplitude-damping channel}, \emph{Chinese Physics} \textbf{16} (2007)
  1843--1847.

\bibitem{Loerincz96}
K.~Loerincz, Z.~Gingl and L.~B. Kiss, \enquote{A stochastic resonator is able
  to greatly improve signal-to-noise ratio}, \emph{Physics Letters A}
  \textbf{224} (1996) 63--67.

\bibitem{Galdi98}
V.~Galdi, V.~Pierro and I.~M. Pinto, \enquote{Evaluation of
  stochastic-resonance-based detectors of weak harmonic signals in additive
  white {G}aussian noise}, \emph{Physical Review E} \textbf{57} (1998)
  6470--6479.

\bibitem{Chapeau99b}
F.~Chapeau-Blondeau, \enquote{Noise-assisted propagation over a nonlinear line
  of threshold elements}, \emph{Electronics Letters} \textbf{35} (1999)
  1055--1056.

\bibitem{Gingl00}
Z.~Gingl, R.~Vajtai and L.~B. Kiss, \enquote{Signal-to-noise ratio gain by
  stochastic resonance in a bistable system}, \emph{Chaos, Solitons and
  Fractals} \textbf{11} (2000) 1929--1932.

\bibitem{Stocks01}
N.~G. Stocks, \enquote{Information transmission in parallel threshold arrays:
  {S}uprathreshold stochastic resonance}, \emph{Physical Review E} \textbf{63}
  (2001) 041114,1--9.

\bibitem{Kish01}
L.~B. Kish, G.~P. Harmer and D.~Abbott, \enquote{Information transfer rate of
  neurons: Stochastic resonance of {S}hannon's information channel capacity},
  \emph{Fluctuation and Noise Letters} \textbf{1} (2001) L13--L19.

\bibitem{Rousseau02}
F.~Chapeau-Blondeau and D.~Rousseau, \enquote{Noise improvements in stochastic
  resonance: {F}rom signal amplification to optimal detection},
  \emph{Fluctuation and Noise Letters} \textbf{2} (2002) L221--L233.

\bibitem{McDonnell02b}
M.~D. McDonnell, D.~Abbott and C.~E.~M. Pearce, \enquote{A characterization of
  suprathreshold stochastic resonance in an array of comparators by correlation
  coefficient}, \emph{Fluctuation and Noise Letters} \textbf{2} (2002)
  L205--L220.

\bibitem{McDonnell08}
M.~D. McDonnell, N.~G. Stocks, C.~E.~M. Pearce and D.~Abbott, \emph{Stochastic
  Resonance: From Suprathreshold Stochastic Resonance to Stochastic Signal
  Quantization} (Cambridge University Press, Cambridge, 2008).

\bibitem{Patel09}
A.~Patel and B.~Kosko, \enquote{Optimal noise benefits in {N}eyman-{P}earson
  and inequality-constrained statistical signal detection}, \emph{IEEE
  Transactions on Signal Processing} \textbf{57} (2009) 1655--1669.

\bibitem{Delahaies13}
A.~Delahaies, F.~Chapeau-Blondeau, D.~Rousseau and F.~Franconi, \enquote{Tuning
  the noise in magnetic resonance imaging to maximize nonlinear information
  transmission}, \emph{Fluctuation and Noise Letters} \textbf{12} (2013)
  1350005,1--16.

\bibitem{Ting99}
J.~J.~L. Ting, \enquote{Stochastic resonance for quantum channels},
  \emph{Physical Review E} \textbf{59} (1999) 2801--2803.

\bibitem{Bowen06}
G.~Bowen and S.~Mancini, \enquote{Stochastic resonance effects in quantum
  channels}, \emph{Physics Letters A} \textbf{352} (2006) 272--275.

\bibitem{Wilde09k}
M.~M. Wilde and B.~Kosko, \enquote{Quantum forbidden-interval theorems for
  stochastic resonance}, \emph{Journal of Physics A} \textbf{42} (2009)
  465309,1--23.

\bibitem{Caruso10}
F.~Caruso, S.~F. Huelga and M.~B. Plenio, \enquote{Noise-enhanced classical and
  quantum capacities in communication networks}, \emph{Physical Review Letters}
  \textbf{105} (2010) 190501,1--4.

\bibitem{Lee11}
C.~K. Lee, L.~C. Kwek and J.~Cao, \enquote{Stochastic resonance of quantum
  discord}, \emph{Physical Review A} \textbf{84} (2011) 062113,1--5.

\bibitem{Lupo13}
C.~Lupo, S.~Mancini and M.~M. Wilde, \enquote{Stochastic resonance in
  {G}aussian quantum channels}, \emph{Journal of Physics A} \textbf{46} (2013)
  045306,1--15.

\bibitem{Gillard19}
N.~Gillard, E.~Belin and F.~Chapeau-Blondeau, \enquote{Stochastic resonance
  with unital quantum noise}, \emph{Fluctuation and Noise Letters} \textbf{18}
  (2019) 1950015,1--15.

\bibitem{Chapeau15c}
F.~Chapeau-Blondeau, \enquote{Qubit state estimation and enhancement by quantum
  thermal noise}, \emph{Electronics Letters} \textbf{51} (2015) 1673--1675.

\bibitem{Gillard17}
N.~Gillard, E.~Belin and F.~Chapeau-Blondeau, \enquote{Stochastic antiresonance
  in qubit phase estimation with quantum thermal noise}, \emph{Physics Letters
  A} \textbf{381} (2017) 2621--2628.

\bibitem{Gillard18}
N.~Gillard, E.~Belin and F.~Chapeau-Blondeau, \enquote{Qubit state detection
  and enhancement by quantum thermal noise}, \emph{Electronics Letters}
  \textbf{54} (2018) 38--39.

\bibitem{Gillard18b}
N.~Gillard, E.~Belin and F.~Chapeau-Blondeau, \enquote{Enhancing qubit
  information with quantum thermal noise}, \emph{Physica A} \textbf{507} (2018)
  219--230.

\end{thebibliography}

\end{document}